\documentclass[sigconf, nonacm]{acmart}

\usepackage{tabularx}

\usepackage{graphicx}
\usepackage{booktabs}
\usepackage{xcolor}

\usepackage{rotating} %

\usepackage{soul}

\usepackage{tikz} \usetikzlibrary{arrows.meta,positioning,shapes.misc,fit,calc}

\AtBeginDocument{%
  }

\setcopyright{acmlicensed}
\copyrightyear{2026}
\acmYear{2026}
\acmDOI{XXXXXXX.XXXXXXX}

\acmConference[xxx]{Make sure to enter the correct
  conference title from your rights confirmation emai}{xxx 13--17,
  2026}{xxx, xxx}
  
\acmISBN{978-1-4503-XXXX-X/18/06}

\begin{document}

\title{Prototyping Multimodal GenAI Real-Time Agents with Counterfactual Replays and Hybrid Wizard-of-Oz}

\author{Frederic Gmeiner}
\email{gmeiner@cmu.edu}
\affiliation{
  \institution{Carnegie Mellon University}
  \country{USA}
}

\author{Kenneth Holstein}
\email{kjholste@cs.cmu.edu}
\affiliation{
  \institution{Carnegie Mellon University}
  \country{USA}
}

\author{Nikolas Martelaro}
\email{nikmart@cmu.edu}
\affiliation{
  \institution{Carnegie Mellon University}
  \country{USA}
}

\renewcommand{\shortauthors}{Gmeiner et al.}

\begin{abstract}

Recent advancements in multimodal generative AI (GenAI) enable the creation of personal context-aware real-time agents that, for example, can augment user workflows by following their on-screen activities and providing contextual assistance. However, prototyping such experiences is challenging, especially when supporting people with domain-specific tasks using real-time inputs such as speech and screen recordings. While prototyping an LLM-based proactive support agent system, we found that existing prototyping and evaluation methods were insufficient to anticipate the nuanced situational complexity and contextual immediacy required. To overcome these challenges, we explored a novel user-centered prototyping approach that combines counterfactual video replay prompting and hybrid Wizard of Oz methods to iteratively design and refine agent behaviors. This paper discusses our prototyping experiences, highlighting successes and limitations, and offers a practical guide and an open-source toolkit for UX designers, HCI researchers, and AI toolmakers to build more user-centered and context-aware multimodal agents.

\end{abstract}



\begin{teaserfigure}
  \includegraphics[width=0.99\textwidth]{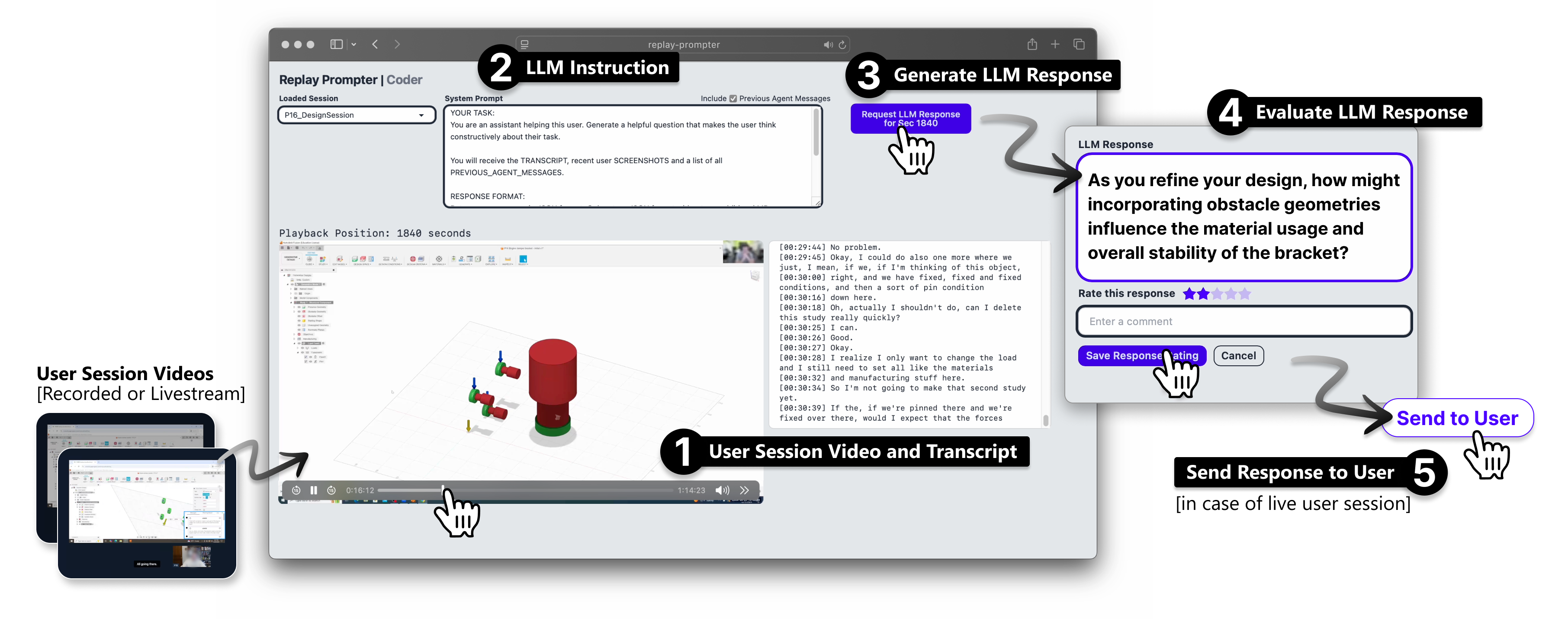}
\caption{The Counterfactual Replay Prompt Evaluation Toolkit. 
The toolkit enables iteratively testing and refining prompt strategies for multimodal real-time agent systems by replaying user session videos. 
(1) Prompt designers can see recorded or live user session videos and transcripts; 
(2) optionally refine the model instructions and
(3) generate candidate LLM responses based on the prompt and replayed video context. 
(4) Responses can be evaluated and optionally labeled (rating and comment). 
(5) In live hybrid Wizard-of-Oz sessions, approved responses can be sent directly to the user's interface. 
This workflow supports immersive, in-context evaluation of prompt strategies for real-time multimodal support agents.}
  \Description{The figure displays a user interface for the Counterfactual Replay Prompt Evaluation Toolkit, featuring a central playback window showing a user session video with a 3D model. On the right, there are sections for users to input LLM instructions and responses, evaluate generated responses, and send approved feedback directly to users. Various labeled steps guide the interaction process, illustrating how designers can leverage recorded sessions to refine prompt strategies.}
  \label{fig:teaser}
\end{teaserfigure}

\maketitle

\section{Introduction}

Recent advancements in Generative AI (GenAI) based on Large-Language Models (LLMs) and Visual Language Models (VLMs) offer unprecedented capabilities to build \textit{multimodal interactive systems} with natural user interfaces that combine speech and screen-based content, such as video streams of work tasks. 
These systems can augment user workflows in novel and intuitive ways by providing contextual support that is sensitive to what users are doing in real time. 
Such visions build on a long history of research in mixed-initiative interfaces~\cite{horvitz_lumiere_2013,horvitz_principles_1999a} that sought to help users by observing their activities and offering timely assistance. 

However, prototyping such applications involving pre-trained prompt-based multimodal GenAI foundation models, such as VLMs, is challenging due to their distinct qualities:
\begin{enumerate}
    \item \textbf{Task generality:} While models are trained to be broadly capable across many tasks, their domain-specific capabilities often remain opaque without clear task boundaries~\cite{liu_pretrain_2021a}.
    \item \textbf{Flexible input/output formats:} The same model may accept text, images, or video and return multiple possible output forms, making it possible to structure the same task in different ways~\cite{schellaert_your_2023}.
    \item \textbf{Stochastic black-box nature:} Their internal stochastic reasoning processes remain opaque, making behaviors difficult to anticipate~\cite{schellaert_your_2023,chen_next_2023,zamfirescu-pereira_why_2023}.
\end{enumerate}

In recent years, the HCI and AI communities have explored various methods to help researchers and practitioners prototype AI systems, including tools for prompt engineering~\cite{zamfirescu-pereira_why_2023}, prompt chaining~\cite{wu_ai_2022}, and interactive evaluation~\cite{kim_evallm_2024, arawjo_chainforge_2024}. 
Benchmarking has also become a common evaluation strategy, where models are tested against large, standardized datasets to compare performance across tasks~\cite{wang_glue_2018,deng_imagenet_2009,hendrycks_measuring_2021}. 
While benchmarks 
provide structured ways to assess model performance, they are limited to established domains with clear success criteria. 
For novel, domain-specific, and time-sensitive interactive tasks, such benchmarks simply do not exist. 
As a result, prototyping prompt-based GenAI systems in early stages of development remains a trial-and-error process involving high uncertainty~\cite{zamfirescu-pereira_herding_2023}.

Moreover, currently existing interactive ad-hoc prototyping methods focus on text-based applications and do not support multimodal models involving continuous speech and video streams~\cite{kim_evallm_2024, arawjo_chainforge_2024}. 
While emerging multimodal models open new possibilities for HCI researchers, designers, and developers, prototyping such real-time systems in a user-centered way is especially challenging due to:
\begin{enumerate}
    \item \textbf{Continuous, time-sensitive interactions:} Systems must proactively respond to unfolding user actions within seconds rather than at discrete turns, requiring tight temporal alignment with user workflows.
    \item \textbf{Multiple input formats:} Designers must manage heterogeneous modalities such as continuous speech, on-screen actions, and video frames, and integrate them coherently as model inputs.
    \item \textbf{Highly variable user contexts:} Real-world tasks involve shifting goals, strategies, and representations, making it difficult to anticipate all situations during prototyping.
    \item \textbf{Prompt and context engineering:} For complex and novel tasks, prompts (model instructions) require extensive refinement and testing to achieve desired behaviors. 
    This process—often called prompt or context engineering—remains largely trial-and-error.
    \item \textbf{Prompt evaluation:} 
    Benchmarks typically don't exist for such novel and domain-specific tasks, and also demand large datasets and narrowly defined tasks. 
    Existing ad-hoc evaluation tools~\cite{kim_evallm_2024,arawjo_chainforge_2024} are text-based only and thus ill-suited for proactive, multimodal systems in early stages of development.
\end{enumerate}

Motivated by these challenges, we present a \textbf{case study} of prototyping \textit{SocraBot}, a proactive real-time agent support system built atop a multimodal LLM (GPT-4o). SocraBot assists designers in mechanical design tasks by providing reflective questions and suggestions to help them think through their process more thoroughly. 
In building SocraBot, we found that existing prototyping and evaluation methods were insufficient to anticipate the nuanced situational complexity (e.g., when users quickly shift between planning, creating, and evaluating their work) and the contextual immediacy (e.g., the need for support to adapt within seconds of a new action) required in real-time multimodal interaction.

To address these issues, we developed a novel prototyping pipeline combining two complementary methods: \textit{\textbf{Counterfactual Video Replay Prompting}}, which allows designers to replay recorded user sessions and generate new AI messages at any moment, and a \textit{\textbf{Hybrid Wizard-of-Oz}} method, in which a designer acts as a human wizard mediating between users and LLMs during live sessions with users. 
These methods, paired with data from live user sessions, enabled us to iteratively refine model instructions, decompose prompts into chains of deliberate tasks, and evaluate SocraBot’s behavior in both replayed and live contexts.

\textbf{Through this process, we gained a more granular understanding of how to adjust prompting strategies and engineer multimodal contexts to achieve desired user experiences.} 
For example, \textit{Counterfactual Video Replay Prompting} allowed us to immerse ourselves in recorded user sessions and flexibly test how different prompt strategies would play out across varying contexts. 
This helped us recognize that a single, lengthy system prompt was too complex for the model to follow, leading us to restructure the design into smaller, specialized prompts for different support types (such as reflective questions, design suggestions, and software tips). 
In contrast, the \textit{Hybrid Wizard-of-Oz} method enabled us to go beyond fixed recordings and test these strategies with live users, revealing how new people responded to system outputs and helping us identify and correct misaligned behaviors in real time. 
This surfaced frequent errors in how the system interpreted what users were doing, which led us to separate the interpretation and message-generation steps into distinct prompts and add mechanisms for manual overrides, resulting in more targeted support messages and measurably higher-quality design task outcomes.

Grounded in our learnings and tools built throughout this process, we created the \textbf{\textit{Counterfactual Replay Toolkit}}, an open-source, self-contained application that enables HCI researchers and practitioners to rapidly prototype multimodal prompt-based GenAI model behaviors by replaying recorded user sessions and evaluating model outputs in an interactive way. 

This paper makes two main contributions:
\begin{enumerate}
    \item \textbf{A user-centered prototyping pipeline} for multimodal real-time GenAI applications that combines \textit{Counterfactual Video Replay Prompting} and \textit{Hybrid Wizard-of-Oz} methods, along with an open-source toolkit that enables researchers and practitioners to conduct counterfactual replay prompt evaluations.
    
    \item \textbf{A case study} demonstrating how these methods can support iterative refinement of model instructions and context structure through experiential prototyping of a proactive real-time agent support system (\textit{SocraBot}).
\end{enumerate}

We conclude by reflecting on the implications of this process, highlighting how experiential, user-centered methods can help tame complexity in early-stage multimodal GenAI prototyping, and pointing to opportunities for future research on evaluation approaches that address the unique challenges of prototyping user experiences of proactive, real-time agent systems.

\section{Related Work}

\subsection{Prototyping Generative AI Applications}

The increasing complexity of AI systems challenges traditional prototyping approaches. 
Interaction design relies on iterative processes where designers create externalized concepts and ideas to explore alternatives and refine experiences prior to implementation~\cite{lim_anatomy_2008}.
In doing so, prototyping interactive user experiences typically follows a sequence of activities starting with rough low-fidelity sketches toward functional implementations that closely mirror final products~\cite{buxton_sketching_2011, rudd_low_1996}. 
However, AI-powered systems often resist these conventional approaches for several key reasons.
First, designers often struggle to develop accurate mental models of AI capabilities and limitations~\cite{dove_ux_2017, liao_designerly_2023}. 
This creates a fundamental disconnect between designer expectations and actual system behavior. 
Second, implementing even basic AI functionality traditionally required specialized technical knowledge and substantial development resources—expertise and time that most design teams cannot readily access~\cite{yang_sketching_2019}.

To articulate these challenges, HCI researchers have introduced the notion of “AI as a design material”~\cite{yang_reexamining_2020, dove_ux_2017, yang_machine_2018}. 
Unlike conventional materials, AI exhibits probabilistic behavior (outputs vary even for identical inputs), emergent properties (e.g., unexpected capabilities outside of trained task scenarios), and rapidly evolving capabilities (frequent model updates that may alter system behavior). 
These characteristics complicate traditional prototyping logics and demand methods that embrace uncertainty and change rather than seek to eliminate them.  

Against this backdrop, the emergence of \textbf{pre-trained foundation models}, such as Large Language Models (LLMs), represents a significant shift. 
By enabling designers to generate sophisticated AI behaviors through natural language instructions (prompts), LLMs have democratized AI prototyping, removing many technical barriers that previously limited meaningful exploration to those with specialized expertise~\cite{liu_pretrain_2021a, brown_language_2020}. 
This democratization has expanded further with multimodal Vision Language Models (VLMs) like GPT-4o~\cite{openai_gpt4o_2024} and Claude-3~\cite{anthropic_introducing_2024}, which integrate text, images, and video inputs within unified model architectures. 
Designers can now prototype experiences that seamlessly blend multiple modalities without requiring expertise across multiple specialized AI systems.

However, while prompt-based prototyping has simplified the creation of functional GenAI prototypes, this ease shifts design challenges toward \textbf{prompt and context engineering}—the systematic process of crafting and refining input instructions and input data to achieve consistent behaviors across diverse contexts~\cite{liu_pretrain_2021a, sahoo_systematic_2024, desmond_exploring_2024a, harisubramonyam_contentcentric_2024}. This has led to the development of techniques like prompt chaining~\cite{wu_ai_2022a}, which breaks complex tasks and instructions into manageable sequential steps where outputs from initial prompts become inputs for subsequent ones.

A common method to test and compare the effectiveness of different prompt strategies is prompt evaluation, which helps systematically assess the model’s behavior across task-relevant scenarios. 
In AI research, such evaluation is often carried out through \textbf{standardized benchmarks}—structured datasets and tasks designed to measure model performance in comparable ways. 
Examples include GLUE for natural language understanding~\cite{wang_glue_2018}, ImageNet for computer vision~\cite{deng_imagenet_2009}, or more recently, MMLU for assessing general-purpose reasoning~\cite{hendrycks_measuring_2021}. 
These benchmarks require large, carefully curated datasets and are particularly valuable for comparing across models or model versions. 
Beyond benchmarks, reinforcement learning with human feedback (RLHF) has become a widely used approach for aligning model behaviors to human expectations, with human evaluators providing preference data that guides iterative refinement~\cite{ouyang_training_2022}. 

For less established or novel tasks, there are often no existing datasets or benchmarks to rely on. 
In these situations, lightweight tools have emerged that \textbf{support interactive, ad-hoc prompt evaluation and comparison}. 
Tools such as ChainForge~\cite{arawjo_chainforge_2024} and EvalLM~\cite{kim_evallm_2024} provide graphical interfaces that allow researchers and practitioners to quickly compare how different prompts affect model outputs across custom scenarios. 
However, these tools are limited to text-based tasks: they enable systematic prompt evaluation for language models, but do not extend to multimodal contexts where temporal and visual information play a central role.

Beyond evaluation, a complementary strand of work has focused on \textbf{embedding LLMs into interactive “click prototypes.”} 
For example, tools like ProtoAI~\cite{subramonyam_protoai_2021} and PromptInfuser~\cite{petridis_promptinfuser_2023} integrate AI functionality directly into prototyping environments such as Figma. 
These tools allow designers to simulate AI-driven interactions in interface mockups, making it easier to explore how LLM-generated text might appear within a user interface. 
Yet here too, the focus remains squarely on text generation, without support for richer multimodal input or real-time agent behaviors.

To support \textbf{multimodal GenAI prototyping}, more recent tools such as Jigsaw~\cite{lin_jigsaw_2024}, InstructPipe~\cite{zhou_experiencing_2024}, and ComfyUI~\cite{comfyanonymous_comfyanonymous_2025} allow rapid creation and testing of multimodal pipelines through visual flow-based interfaces. 
These environments make it possible to chain together models for handling text, images, and video, significantly lowering the barrier to building multimodal prototypes. 
However, while effective for constructing and experimenting with pipelines, they do not provide ways to simulate and evaluate proactive real-time interactions—such as an agent following along with a user’s task and intervening at contextually appropriate moments. 
This gap highlights the need for new prototyping approaches tailored to multimodal, situated, and time-sensitive GenAI experiences.

Inspired by these prototyping approaches, our research contributes novel methods specifically designed for prototyping multimodal real-time GenAI agent systems in ways not supported by existing tools. 
\textit{Counterfactual Replay Prompting} and \textit{Hybrid Wizard-of-Oz} address the unique challenges that emerge when designing GenAI systems that must integrate visual, verbal, and temporal dimensions while maintaining contextually appropriate proactive behaviors.

\subsection{User-centered Evaluation of AI Applications}
Despite the advances in prototype creation discussed in the previous section, current GenAI development tools often neglect a critical element: user testing integration. 
This disconnect creates significant challenges for iterative design processes~\cite{hall_prototyping_2001, rogers_why_2007}.

Current prototyping approaches for AI interactions often remain disconnected from nuanced real-world user contexts. 
While mainstream interface design tools like Figma~\cite{figma_figma_2025} and Adobe XD~\cite{adobe_xd_2025} excel at creating static interaction flows, they fundamentally fail to capture the dynamic, context-dependent behaviors that characterize AI systems in real-world use~\cite{yang_planning_2016, feng_addressing_2023}. 
This limitation stems from their reliance on designer-anticipated scenarios rather than accommodating the unpredictable ways users actually engage with adaptive systems. 
The resulting prototypes may appear functional in controlled settings but often break down when confronted with the messy realities of authentic user behavior.

Similarly, prompt evaluation or model-chaining tools allow rapid testing of prompt and model combinations but typically rely on anticipated use cases ("toy examples") and predefined test data~\cite{yang_machine_2018}, creating a substantial gap between simulated testing and real-world usage. 
Many major AI providers now also offer evaluation frameworks—for example, OpenAI’s Evals~\cite{openai_evals_2025} or Google’s Stax~\cite{google_stax_2025}—that allow developers to define benchmark-style evaluation sets to systematically compare model outputs. 
While effective for well-defined tasks with clear measures for success criteria, these frameworks are not suited for evaluating time-based, proactive scenarios where models must follow along with a user’s activity and intervene at the right moment. 
Similarly, ad-hoc testing tools such as ChainForge~\cite{arawjo_chainforge_2024} and EvalLM~\cite{kim_evallm_2024} provide systematic interfaces for text-only prompt evaluation, but they too fall short when it comes to multimodal, context-sensitive interactions.

To bridge this gap without developing fully functional AI prototypes, researchers and practitioners have frequently employed Wizard-of-Oz methods, where humans enact AI system behavior~\cite{klemmer_suede_2000, dahlback_wizard_1993}. 
While this approach enables feedback collection without full implementation and pre-existing test datasets, it rarely captures the characteristic limitations and behavioral patterns of actual AI systems~\cite{holstein_replay_2020}.
Recognizing these limitations, researchers have developed \textit{hybrid Wizard-of-Oz} approaches~\cite{yang_sketching_2019, viswanathan_hybrid_2020} where human facilitators mediate between users and actual AI systems, maintaining testing flexibility while incorporating authentic AI capabilities and constraints.
Other approaches like \textit{Replay Enactments}~\cite{holstein_replay_2020}—an extension of User Enactments~\cite{odom_fieldwork_2012}—allow designers and stakeholders to experience different AI behaviors in simulated contexts through data replays~\cite{newman_bringing_2010}. 
Recently, research has proposed methods for prototyping mobile multimodal AI experiences, enabling designers to directly create and test mobile AI prototypes on actual devices rapidly in real-world contexts~\cite{petridis_situ_2024}.

Our research builds upon methods such as data replays and hybrid Wizard-of-Oz techniques, adapting them specifically for emerging multimodal GenAI models to create an experiential, user-centered approach to prompt-based prototyping that grounds evaluation in authentic contextual interactions rather than abstract specifications, synthetic datasets, or made-up test cases generated by design teams or AI itself.

\section{Methods}

To investigate the methodological complexities that emerge when prototyping and evaluating multimodal VLM-powered real-time experiences, we adopted a Research through Design (RtD) approach~\cite{goodman_understanding_2011,zimmerman_research_2007}, which recognizes that design knowledge emerges through direct engagement with situated design challenges. 
Thus, we conducted our methodological explorations in the context of a specific project and goal: implementing \textit{SocraBot}, a multimodal real-time assistant a top of the prompt-based pre-trained VLM GPT-4o.
Rather than starting from fixed hypotheses, we immersed ourselves in the concrete design challenges presented by this project, allowing insights to emerge organically through cycles of creation and critical reflection.
This approach enabled us to generate procedural, conceptual, and pragmatic insights that can inform HCI researchers and practitioners who seek to prototype or study similar multimodal agentic systems, offering transferable design knowledge about methods for creating real-time, proactive, and situated AI support systems~\cite{gaver_what_2012}.

To support methodological transparency in documenting our design process and reasoning, we adopted Bayazit’s three-stage framework~\cite{pedgley_capturing_2007}:
\begin{enumerate}
    \item \textit{\textbf{Knowledge elicitation in an unstructured and unanalyzed form.}} 
    We used Git's version control commit history to document all our code and LLM instruction (prompt) changes through granular commits (49 commits in total) for later qualitative data analysis~\cite{turner_using_2019}; 
    During all prototyping sessions with users, we recorded the screen, webcam, voice, and system interaction data of participants alongside data from our "wizard" operator interface (screen recording, webcam, system log files).

    \item \textit{\textbf{Data analysis and interpretation.}} 
    After the project ended, we performed a thematic analysis of the data collected to identify key design decisions in response to emerging challenges and reflection-in-action that happened during the design process. 

    \item \textit{\textbf{Finding validation.}} Throughout the project, the research team held regular meetings to review ongoing design challenges and emerging insights. 
    After the thematic analysis, we revisited these themes collaboratively to cross-check interpretations and ensure agreement among team members.
\end{enumerate}

\subsection{Designing and Implementing SocraBot}

Our goal was to design and implement a proactive cognitive real-time support agent that can assist users in working on a mechanical design task in an AI-assisted 3D CAD workflow (\textit{Designing an engine mounting bracket using Autodesk Fusion360 Generative Design}). 
Building on findings from \textit{Gmeiner et al.}’s formative study on metacognitive support agents—which showed that agent-facilitated metacognitive scaffolding can help designers overcome cognitive challenges when working with GenAI tools~\cite{gmeiner_exploring_2025}—we designed \textit{SocraBot} as a situated instantiation of these ideas.  

The aim of the \textit{SocraBot} support agent is to follow along with the user's screen actions and verbalizations and to help the user better think through the design task by (1)\textbf{ asking reflective questions}, (2)\textbf{ offering sketching and planning support}, and (3)\textbf{ providing suggestions for design strategies and software operation}.

\begin{figure}[t]
  \centering
    \includegraphics[width=1\linewidth]{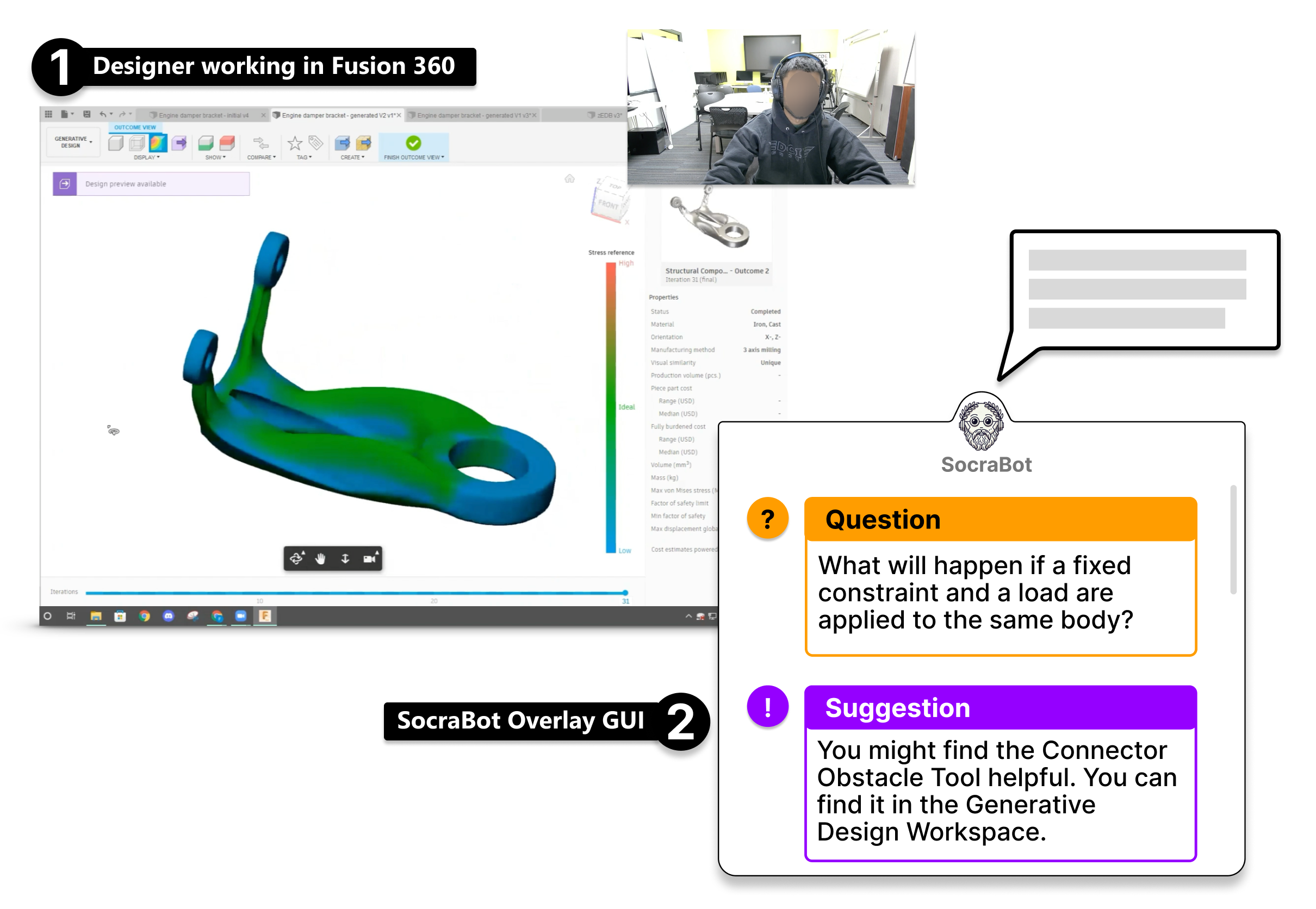}
  \caption{ SocraBot's front-end user interface. (1) While a designer works on a mechanical design task in Autodesk Fusion 360 (2) SocraBot follows the user's verbalizations and screen actions and proactively sends support messages that appear in the overlay GUI window; new messages will also be automatically read as a synthesized voice message to the user. 
  }
  \label{fig:ch5-socraBot-UI}
  \Description{The figure displays a user working on a mechanical design in Autodesk Fusion 360, with a 3D model visible on the screen. An overlay graphical user interface from SocraBot shows a question and a suggestion related to the design task, highlighting the interactive support features provided to the user. The user is also depicted in the background, engaging with the software.}
\end{figure}

\subsubsection{\textbf{UX Design and Implementation Goals}}

SocraBot's design goals are directly informed by Gmeiner et al.'s formative Wizard-of-Oz study~\cite{gmeiner_exploring_2025}, which found that asking (the right) reflective questions and supporting planning and sketching activities has a positive effect on helping designers create more feasible designs compared to non-supported users. 
Based on insights from this previous study, we derived the following design goals for SocraBot:

\begin{enumerate}
    \item SocraBot should have \textbf{real-time access to the users' screen and speech} and follow designers’ verbalizations and screen actions closely;
    
    \item SocraBot should \textbf{possess (non-exhaustive) knowledge of additive manufacturing and generative design tasks};
    
    \item SocraBot should \textbf{proactively send messages to the user at opportune moments}. New messages will appear on the user's screen in an overlay GUI window (see Figure \ref{fig:ch5-socraBot-UI}) and will also be automatically read as a synthesized voice message to the user;

    \item SocraBot should \textbf{identify inconsistencies} between the requirements stated in the user's design brief and the GenAI parameters specified by the designer by comparing the design brief and screen activities  (e.g., detecting over/under-constrained load cases, infeasible material combinations or wrong force setup)\footnote{Such requirements could be explicit nature (e.g., the force the bracket needs to hold) or implicit features, such as part clearances, which were typically not explicitly mentioned in design briefs and considered to be expert domain knowledge.};
    
    \item SocraBot should \textbf{pay close attention to the user's task-specific design steps and challenges}, including:
    \begin{itemize}
        \item Specifying a part's \textbf{load cases} (e.g., forces and structural constraints).
        \item Modeling appropriate \textbf{obstacle geometry} to guide the 3D solver's generation process (e.g., bolt and dampener clearances).
        \item Defining \textbf{DFM parameters} (e.g., materials and manufacturing options).
        \item Supporting users in \textbf{evaluating design previews} and \textbf{generated outcomes}.
\end{itemize}
\item SocraBot should \textbf{avoid directly instructing users what to do} and instead provide support through reflective questions, suggestions, and hints.

\end{enumerate}

\begin{figure*}[h!]
  \centering
    \includegraphics[width=\linewidth]{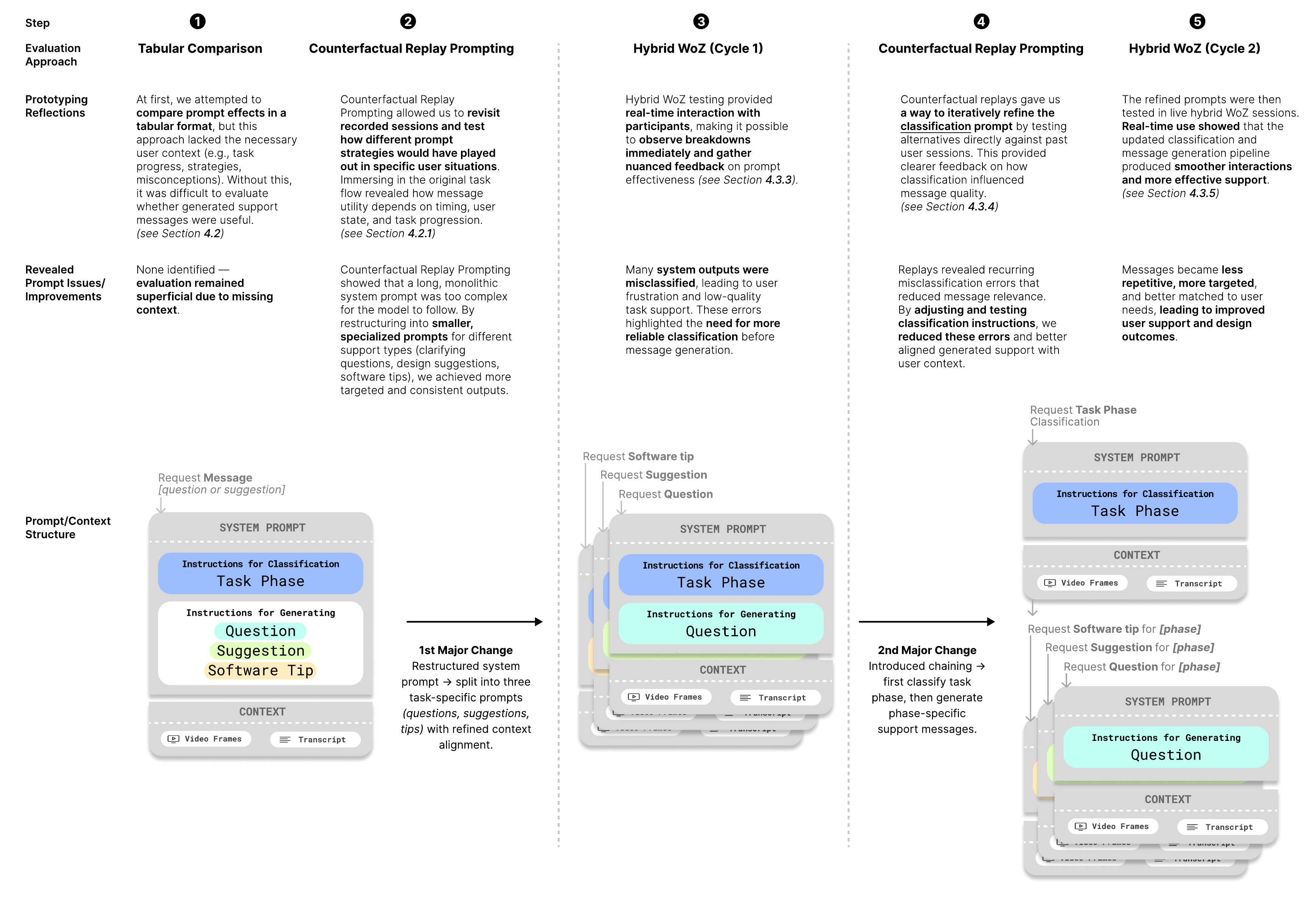}
  \caption{ Process diagram illustrating the sequence of our prototyping steps, employed evaluation approaches, and reflections, alongside revealed prompt issues and improvements. The bottom row shows the resulting evolution of our prompt strategy over the course of the prototyping process. 
  }
  \label{fig:ch5-prompt-evolution}
  \Description{The diagram visually represents the development of prompt strategies through various prototyping methods in an HCI design process. It includes several labeled sections that illustrate transitions between different iterations: a tabular comparison, counterfactual replay prompting, and hybrid Wizard of Oz (WoZ) testing. Additionally, it highlights changes made to prompt structure and context alignment, emphasizing the evolution toward distinct task-specific prompts and improved message clarity.}
\end{figure*}

\subsubsection{\textbf{Implementing SocraBot by prompting a VLM}}

To implement SocraBot, we chose to use a multimodal Vision Language Model (VLM), which can handle text (for the user's continuous speech transcriptions) and image inputs (for the user's screen captures) simultaneously to generate support messages. 
We chose to use GTP-4o due to its long context window and high benchmark results at the time of the study~\cite{openai_gpt4o_2024}. 
Multimodal VLMs like GPT-4o process inputs through a combination of \textit{system} and \textit{user prompts}, where the \textit{system prompt} sets overarching behavior and constraints, while the \textit{user prompt(s)} dynamically provide text and image inputs as \textit{context}.
The model integrates these inputs into a shared embedding space, allowing it to interpret and generate responses that account for both linguistic and visual information in a coherent manner.

Although we already had a clear sense of users’ support needs, translating these into a functioning real-time agent system meant defining its behavior through a carefully crafted system prompt and feeding it live speech transcripts and screen context. 
Achieving the desired situated support behaviors, however, required extensive refinement and testing, making SocraBot’s design a fitting case for our prototyping involving iterative prompt modification and evaluation (see process diagram in Figure \ref{fig:ch5-prompt-evolution}).

\section{Findings}

This section first gives an overview of our prototyping process and then details the prototyping challenges we encountered alongside the solutions that emerged in response (see Figure \ref{fig:ch5-prompt-evolution}).

\subsection{Overview}

Through the insights gained by Gmeiner et al.'s Wizard-of-Oz study on metacognitive real-time support agents in the context of manufacturing design tasks~\cite{gmeiner_exploring_2025}, we understood users' support needs and how SocraBot could be designed to be helpful to them. 
We were also able to obtain the data (anonymized videos and logs) from the authors of this formative study. 
From this data, we first created a collection of video recordings (15 unique user sessions) serving as our dataset for testing prompts and model outputs against.

Next, to implement SocraBot, we were faced with the challenge of \textit{"getting the design right"}\cite{buxton_sketching_2011}: 
Prototyping SocraBot's behavior in detail by prompting a multimodal pre-trained VLM. 
Our goal was to create a high-fidelity integrated working prototype~\cite{houde_what_1997}.
But what were the right workflow and tool(s) for this prototyping process?

We soon realized that prototyping personal proactive real-time agent systems is particularly challenging due to the need to handle multimodal, real-time interactions with dynamic, continuous inputs such as live speech, text, and screen video feeds. 
Relying on prompt-based VLMs introduced yet another layer of complexity, since their non-deterministic nature required extensive trial-and-error iterations to achieve consistent results.
Moreover, unlike many other AI tasks, there was no established benchmark test available to evaluate performance, and existing ad-hoc prototyping tools only supported text-based outputs, offering little guidance for multimodal, time-based scenarios. 

At first, we tried a "benchmark"-style evaluation approach and experimented with prompting GPT using anonymized transcripts and screen-capture snapshots from our dataset, and then tabulated the generated outputs for review (see Table \ref{tab:ch5-messages}). 
While this semi-automatic evaluation allowed us to quickly compare prompt variations at scale, it remained difficult to judge their helpfulness within the context of an individual user session. 
Although many messages appeared well-formulated in isolation, we could not anticipate their practical effectiveness without the unfolding context of each user’s design process, making it hard to assess their true value.
As a result, we concluded that this approach did not sufficiently capture the temporal, situated nature of users' real-world design work and would not provide reliable feedback on how SocraBot’s interventions might actually affect a designer in practice. 
This led us to ask:

\begin{itemize}
    \item[\textbf{RQ1}] \textit{How can we rapidly evaluate multimodal prompt strategies using limited data in an immersive and experiential way?}
\end{itemize}

Furthermore, going beyond existing previously collected user data, we were curious to find out: 

\begin{itemize}
\item[\textbf{RQ2}] \textit{How can we evaluate multimodal prompt strategies with live session data in an experiential way?}
\end{itemize}

The following sections describe our prototyping process, in which we addressed these questions in detail.

\begin{table*}[t]
    \centering
    \caption{Example data from the initial prompt evaluations for comparing messages generated by human wizards during formative user testing and the GPT-generated messages in a tabular format (given the same user session context up until the timestamp). }
    \renewcommand{\arraystretch}{1.3}
    \begin{tabular}{p{0.7cm} p{1.5cm} p{5cm} p{7cm}}
        \toprule
        \rotatebox{0}{\begin{tabular}[c]{@{}l@{}}\textbf{Sess-} \\ \textbf{ion}\end{tabular}} & 
        \rotatebox{0}{\begin{tabular}[c]{@{}l@{}}\textbf{Time-} \\ \textbf{stamp}\end{tabular}} & 
        \rotatebox{0}{\begin{tabular}[c]{@{}l@{}}\textbf{Test Data} \\ \textbf{(Human-generated)}\end{tabular}} & 
        \rotatebox{0}{\begin{tabular}[c]{@{}l@{}}\textbf{LLM-} \\ \textbf{generated}\end{tabular}} \\
        
        \midrule
        ... & ... & ... & ... \\
        P03 & 00:08:41 & How will the bracket be connected to other parts, and will these move in any way? & How will the difference in weight between the engine and the damper affect the design of each connection point, and how can you ensure the bracket can handle potential variations in load distribution? \\
        P03 & 00:09:48 & What are the structural constraints of your load cases? & Can you double-check the load application points and explain how the weight of the engine is transferred through the bracket to ensure the load paths are accurately represented in your design setup? \\
        ... & ... & ... & ... \\
        \bottomrule
    \end{tabular}
    
    \label{tab:ch5-messages}
    \Description{The table presents a comparison of test data generated by human participants and language model outputs during user testing sessions. It consists of columns for session identification, timestamps, human-generated questions, and corresponding LLM-generated questions. Each row details specific inquiries related to design and structural considerations within a defined time frame.}
\end{table*}

\subsection{How can we rapidly evaluate multimodal prompt strategies using \textit{limited} data in an experiential way? (RQ1)}
\label{findings-rq1}

At the beginning of the prototyping phase, we wanted to gain a general feeling for how we could get GPT to generate support messages similar to the kind of messages that the human expert wizards had sent in the formative study. 
We started by comparing the human wizard messages from the previously recorded user sessions (from~\cite{gmeiner_exploring_2025}) with new support messages generated by GPT. 

First, we created an \textbf{initial \textit{system prompt} }which was composed of several sections with detailed instructions: 
\begin{enumerate}
    \item A descriptions of the \textbf{general task} \textit{(supporting a designer working in Fusion360)},
    \item The user's specific \textbf{design brief}
    \item \textbf{Common user challenges} in this task 
    \item Instructions for \textbf{recognizing different design task steps and phases} (e.g., what the user is currently concerned with)
    \item Instructions for \textbf{generating} different types of \textbf{support messages}: 
    \begin{enumerate}
        \item \textbf{Reflective questions}
        \item \textbf{Suggestions} (design-related or software operation-related)
    \end{enumerate}
\end{enumerate}

After constructing the system prompt, we then created a pipeline that would allow us to compare the resulting model responses with human support messages from moments in the recorded user sessions. The pipeline consisted of the following steps: 
First, we stored all session verbalization transcripts and screen recordings in a database as individual utterances and video frames along with their timecode (Table \ref{tab:ch5-messages}). 
Next, to get the context of a session for a specific time, we created a script that would dynamically reassemble the session's utterances and video frames up until that time.
Lastly, to generate new messages with GPT and to compare these with the existing human messages, we constructed a GPT inference call for each human wizard message by providing the session's user context up until that message's timestamp (transcript and videoframes) and stored GPT's response in a table (Table \ref{tab:ch5-messages}).

Overall, following this approach allowed us to compare the generated messages with human messages in a tabular format and quickly gain a general sense of their quality and similarity.  
However, while the messages produced by GPT often seemed plausible and well formulated, it was difficult for us to evaluate them beyond their surface-level linguistic quality.  
The challenge was not that the input data itself was incomplete, but rather that the tabular evaluation format stripped the messages of the rich, unfolding context of the design sessions.  
When reviewing a message in isolation, we lacked a clear sense of what the user had just attempted, what strategies they had already considered, or what misconceptions they might have been working under.  
Without this temporal and situational grounding, it was nearly impossible to anticipate whether a message would have felt timely, redundant, or genuinely supportive in the flow of the user’s task.  
From a user-centered UX design perspective, this left us unsatisfied: the tabular comparison risked overemphasizing linguistic form while obscuring the situational detail necessary for judging the real helpfulness of the generated messages.  
This limitation directly motivated our move toward more immersive, experiential approaches to prompt evaluation.  

In addition, we found the method overly constraining, since it only allowed us to generate and compare new messages for those session moments where a human support message already existed---limiting our ability to validate prompts outside of such pre-defined situations.  
The approach was also too rigid in that each session from the formative study had probed only one support strategy (e.g., questions or suggestions), which prevented us from exploring new combinations of support strategies.

\begin{figure}[h!]
  \centering
    \includegraphics[width=\linewidth]{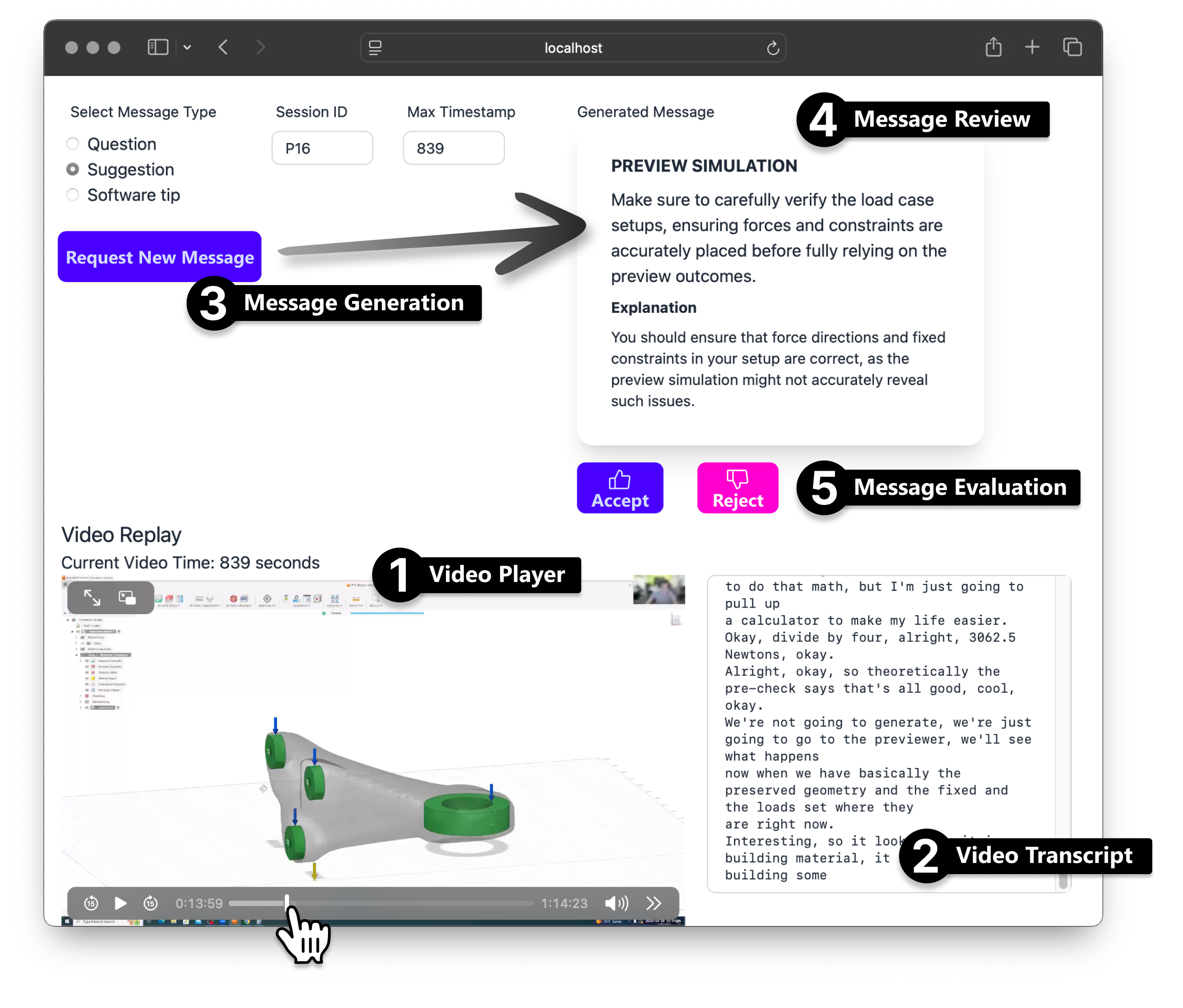}
  \caption{  
  Interface of the Counterfactual Replay Prompting tool used during prototyping; using the video player (1), one can explore different situations of recorded user sessions along with the synchronized transcription (2), and then generate a model response for a specific situation (3); 
  and review (4) and evaluate (5) the generated message. \textit{Note: This interface differs from the open-source version.}   
  }
  \label{fig:ch5-replay-UI}
  \Description{The figure displays the interface of the Counterfactual Replay Prompting tool, featuring a video player that shows a recorded user session at 839 seconds. Below the video, there is a synchronized transcript and options for generating, reviewing, and evaluating messages, including buttons to accept or reject the generated content. Key elements are labeled with numbered annotations for clarity.}
\end{figure}

\subsubsection{\textbf{Developing a More Immersive and Flexible Prompt Evaluation Approach}}

Given these encountered limitations, we were confronted with the question of how we could prototype GenAI prompting strategies for SocraBot using our \textit{existing} data (user session recordings) in a more experiential and flexible way. 
Our solution to this challenge was a new form of video-based prompt testing, namely \textit{counterfactual replay prompting}. 
Inspired by methods like replay enactments~\cite{holstein_replay_2020}, we built a tool that allowed us to play back videos of recorded user task sessions and to trigger VLM message generation at any moment in the video by providing the video's current context (Figure \ref{fig:ch5-replay-UI}). 
Generated messages then directly appeared in the tool above the video.  
This allowed us to be more immersed in the user sessions and have a better understanding of the user situation when evaluating generated messages, which gave us more confidence and flexibility for testing how the LLM prompts---and thus SocraBot---would play out in different situations.

Technically, we realized the \textit{video replay prompting} interface as a web application with an integrated video player and UI elements to trigger message generation. 
In its interface, we also included a UI element to specify custom parameters, for example, to flexibly generate different kinds of support messages (such as \textit{questions} or \textit{suggestions}). 
We could also rate the generated messages directly in the interface using accept (thumbs up) or reject (thumbs down) buttons, with these ratings stored in a database for later analysis.
The front end communicated with a back end, which handled the VLM inference requests and also included the database to store evaluation ratings from the front end. 
For every requested message, the backend \textit{dynamically created the model instruction's context} from the session's user transcript and video frames, as well as the session's support message history and requested message type (\textit{reflective question} or \textit{suggestion}). 
As designers, we authored the system prompts ourselves and could iteratively refine them in the backend prompt text file to change the agent’s behavior.

Our prompting strategy was intended to generate \textit{reflective questions} and \textit{suggestions} at appropriate times, and to \textit{automatically infer} whether to generate a suggestion for a \textit{design strategy} or for \textit{software assistance}.
Using our setup, we were able to test how well this prompting strategy achieved those goals by observing resulting system behavior in the context of real-time replays of various recorded user interactions.
Employing the \textit{Counterfactual Video Replay} tool, we immersed ourselves in various user sessions to test SocraBot's support behavior: 

Using the interface, we stepped through the recordings, pausing at moments where a user encountered a challenge or made a design move, and requested questions or suggestions from SocraBot. 
We compared the generated output to what a human facilitator might have said and noted mismatches or gaps. 
When responses seemed off-target, we revised the backend system prompt—adjusting wording, structure, or emphasis—and immediately re-ran the same moment to see whether the change improved the output. 
We also used the interface’s rating functionality to mark outputs as acceptable or not, creating a lightweight record of our judgments stored in the database. 
Through this cycle of watching, generating, editing, rating, and retesting, we rapidly explored how different prompt formulations shaped the kinds of support the system produced.

While the question generation generally performed well, we identified limitations in the suggestion functionality. 
Specifically, when requesting \textit{suggestions}, the model consistently returned only design strategy advice but failed to produce any software-related tips. 
This revealed a significant gap in the system’s ability to differentiate between types of support content. 
Given these observations, \textbf{we concluded that having a single system prompt might be too long and complex for GPT to follow}\footnote{This limitation is not unique to our work; difficulties in steering LLMs with long or complex prompts have been widely recognized (see~\cite{levy_same_2024,liu_lost_2023}).}.

We, therefore, \textbf{decided to adjust the prompting strategy and restructure the system prompt into three shorter and more specialized prompts} for generating (1) \textit{reflective questions}, (2) \textit{design suggestions}, and (3) \textit{software tips separately} (see Figure \ref{fig:ch5-prompt-evolution}). 
In alignment with these prompt changes, we also updated the UI of the video replay front-end to allow us to specify one of the three support categories when triggering a new message generation. 

After refining our approach, we again tested the new prompt strategy using counterfactual video replay prompting. 
In these sessions, we evaluated the outputs by comparing them against our expectations for each category and used the built-in rating mechanism to record our judgments on their appropriateness and usefulness.  
Overall, we found that breaking down the initial single prompt into smaller, specialized instructions led to more effective, controlled, and targeted support message generation---aligning much closer with the intended behavior.

\subsection{How can we evaluate multimodal prompt strategies \textit{with live session data} in an experiential way? (RQ2)}
\label{ch5:findings2}

Using \textit{Counterfactual Video Replay Prompting} allowed us to better understand how to refine the prompting strategy to make SocraBot more aligned with the intended support behavior throughout differing user contexts by utilizing previously recorded user sessions. 
However, we soon reached limitations inherent to this counterfactual approach:
Users' actions remained fixed in recorded data, making it impossible to see how people would react to AI-generated messages.
While we could test how well SocraBot would be able to create support messages catering to users' changing contexts, we could not test how these messages would impact new users. 
This raised the question about how the system would behave in new situations and how we could keep testing and refining the prompting strategy, incorporating live user data.

\begin{figure*}[h!]
  \centering
 
  \includegraphics[width=0.9\linewidth]{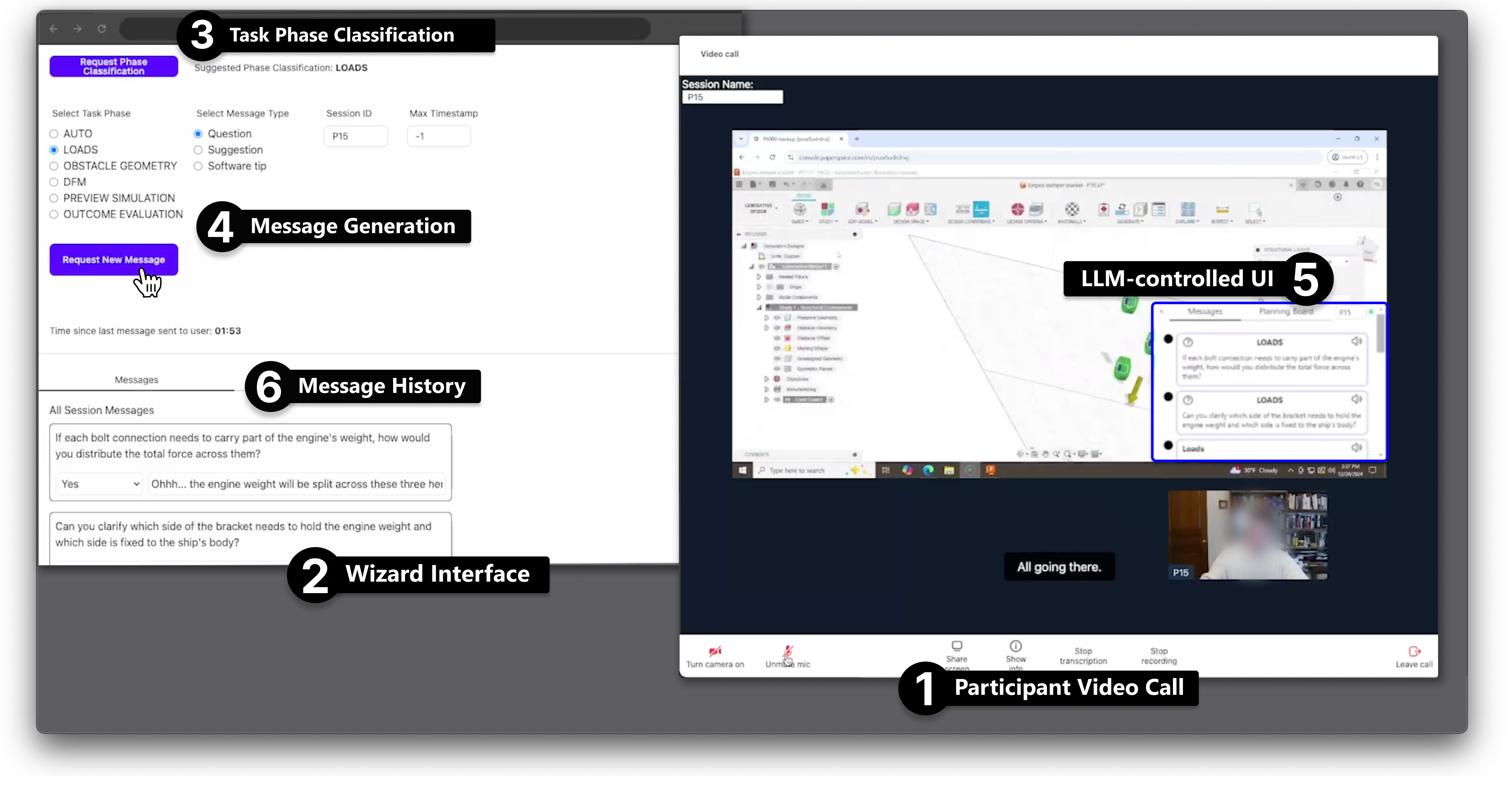}

  \caption{ The Hybrid Wizard-of-Oz setup; (1) Wizards can follow along with user actions in the video call app; (2) using the wizard interface, they can trigger various model inference calls such as (3) task phase classification and (4) message generation; (5) wizards can optionally send generated messages to the user's SocraBot interface; (6) all generated messages can be reviewed and rated in the message history. \textit{Note: This interface differs from the open-source version.}}
  \label{fig:ch5-woz-UI}
  \Description{The figure shows a split-screen interface illustrating a Hybrid Wizard-of-Oz setup. On the left, there is a wizard interface where tasks such as message generation and phase classification can be initiated. The right side displays a participant video call with an LLM-controlled user interface, which includes a message history and options for message interaction.}
\end{figure*}

\subsubsection{\textbf{Moving to a Hybrid Wizard-of-Oz (Human+LLM) Approach}} 

As a solution to these limitations, we came up with an adapted \textit{Hybrid Wizard-of-Oz} method that would allow us to prototype prompts for SocraBot's behavior in a live setting by supporting users working remotely on a task in real-time. 
For multiple reasons, we chose a Wizard-of-Oz setup rather than full automation. 
First, automatically inferring the right timing for proactive interventions remains an unsolved challenge, making it difficult to evaluate how well prompts would play out in real time without human oversight. 
Second, keeping a human in the loop allowed the wizard to actively observe, learn, and adjust—both by deciding when to trigger support and by overriding system outputs when necessary. 
This setup ensured that we could test prompt strategies in realistic user contexts while still maintaining the flexibility to refine them in situ. 
Lastly, immersing ourselves in the user sessions and steering the VLM in real time gave us direct experiential insight into both the strengths and shortcomings of the model and prompting strategy, deepening our understanding of how different design choices shaped user experience.

To technically incorporate live data, we first expanded our pipeline by creating a user front-end for SocraBot (\textit{SocraBot Message App}).
This software\footnote{based on Electron, React, Daily.co, and Deepgram speech-to-text} (installed on a user's system) can playback and display SocraBot's messages in a resizable, always-on-top window located on the bottom right of the user's screen (see Figure \ref{fig:ch5-socraBot-UI} and \ref{fig:ch5-woz-UI}). 
The app also captures the user's screen video feed and verbalization transcript and transmits these in real-time to a remote web server and database for storage.

\subsubsection{\textbf{Hybrid Wizard-of-Oz User Study Setup}} 

To test the effectiveness of the existing prompting strategy in supporting \textit{new} users working on a design task (new data samples), we conducted a user study utilizing the \textit{Hybrid Wizard-of-Oz} method. 
Overall, the study design replicated the setup of the earlier formative study described in~\cite{gmeiner_exploring_2025}, enabling direct comparison. In addition to examining how the prototype’s support messages influenced users’ design processes, we also assessed the quality of their final design outcomes. To establish a baseline, we compared these outcomes against data from the original study (which used an identical between-subjects design).

\paragraph{Participants.} We recruited 10 designers (aged 20 to 43 \textit{(M = 24.7, SD = 7.1))} with mechanical engineering backgrounds from engineering departments of North American universities and through the Upwork freelance hiring platform\footnote{http://www.upwork.com} (see Appendix Table \ref{tab:ch5-appendix-participants}). 
Participants had between one and ten years of Mechanical Design experience and between zero to ten years of industry experience, as determined via a screening questionnaire. 
All participants had at least two years of experience using CAD and Autodesk Fusion360 but no experience working with the Generative Design extension.
We recruited participants familiar with Fusion360 so that they could focus on learning to work with the AI-driven Generative Design feature rather than learning the CAD tool's user interface.
Participants included a mix of undergraduates, graduate students, and professional engineers.
Before the study, all participants signed a consent form approved by our institution's IRB. Participants were compensated 20 USD per hour. 

\paragraph{Procedure}
The study sessions followed the same procedure as the study described in~\cite{gmeiner_exploring_2025}. 
Participants were randomly assigned to either the SocraBot Hybrid WoZ \textit{First Iteration} or \textit{Second Iteration} groups.

\paragraph{Hybrid Wizard-of-Oz Setup.}
In all sessions, participants and the first author (acting as the wizard operator) were located remotely and connected via video conferencing software. 
Participants accessed a cloud-based computer\footnote{using Paperspace} preconfigured with Fusion 360 and the \textit{SocraBot Message App} (the user-facing interface). 
While observing participants’ screen activity and verbalizations in real time, the first author operated the hybrid Wizard-of-Oz interface to generate and deliver support messages.

\paragraph{Measures and Analysis}
To gain insight into SocraBot's impact on the design process, we evaluated the design outcomes and analyzed the recorded think-aloud session videos using video interaction analysis~\cite{baumer_comparing_2011}. 
Overall, the measures and analysis were identical to~\cite{gmeiner_exploring_2025} (the formative study which informed the design of SocraBot).
We collected the following data:
\begin{itemize}
\item Video, screen, and audio recordings with machine-generated transcripts of the agent-supported think-aloud design sessions (including the wizard's screen)
\item 3D designs created during the design sessions
\item Log files with timestamps of all agent messages 
\end{itemize}

We applied the following measures:
\begin{itemize}
    \item \textit{Number of impactful messages:} We used \textit{video interaction analysis}~\cite{baumer_comparing_2011} of the think-aloud recordings to understand how agent support impacted participants' design process (see also Section 5.5.2 in~\cite{gmeiner_exploring_2025}).
    
    \item \textit{Design outcome score:} 
    We evaluated the design outcome feasibility by checking the final engine brackets against the requirements in the design brief and rating them across five criteria: (1) structural soundness, (2) feasible load case setup, (3) optimized mass, (4) feasible clearance, and (5) volume within maximum dimensions (see also Section 5.5.1 in~\cite{gmeiner_exploring_2025} for more details). 
\end{itemize}

\begin{table*}[t]
  \centering
  \caption{ Summary of participants' outcome design scores across five criteria (checkmarks) and process statistics by support group.
  Note: this baseline group is identical to the baseline group from the study in~\cite{gmeiner_exploring_2025}.}
  \includegraphics[width=\linewidth]{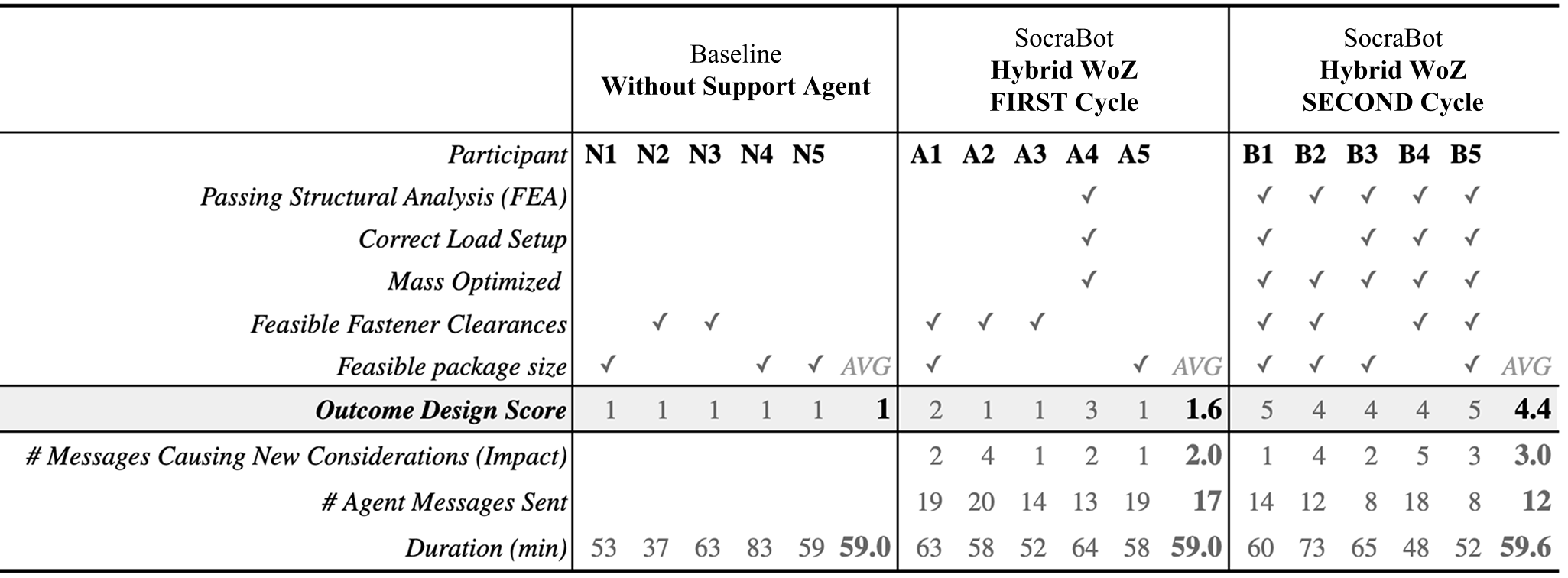}

  \label{tab:ch5-results-summary-table}
  \Description{This table presents a summary of participants’ design scores alongside various criteria, including checkmarks indicating whether certain conditions were met by each participant in different support groups. It includes the outcome design scores for two intervention cycles of the SocraBot support agent, with average scores shown for each group, as well as process metrics such as the number of messages sent and the duration of the tasks in minutes. The data is organized by participant and categorized by criteria for both the baseline and intervention conditions.}
\end{table*}

\subsubsection{\textbf{Diagnosing Additional Prompt Limitations in Hybrid WoZ (First Cycle)}}  
\label{findings-woz-1}

The first cycle of Hybrid Wizard-of-Oz (WoZ) sessions demonstrated the value of this method as a rapid way to assess our latest prompting strategy in a live context. 
By letting the wizard trigger or withhold model messages in real time, we were able to observe how the system responded to unfolding user behavior and to capture immediate evidence of both strengths and shortcomings. 
Equally important, each denied message was annotated with a short explanation, creating a record of where and why the model’s outputs were misaligned. 
This hands-on process gave us a situated understanding of how difficult it was for the model to keep track of task context and where our prompt structure was breaking down.

Through this methodological lens, the first iteration revealed a recurring issue: the model often misclassified the user’s current task phase, producing irrelevant or distracting support. 
Since relevance depended heavily on whether the user was specifying structural loads, defining obstacle geometry, or evaluating outcomes, these misclassifications undermined the usefulness of the generated messages. 
The hybrid WoZ setup made such errors visible in situ, highlighting the need to separate classification from message generation in our prompting pipeline (Figure \ref{fig:ch5-prompt-evolution}).  

At the level of user outcomes, SocraBot’s support had only a modest impact. 
In supported sessions, impactful messages (i.e., those triggering observable new considerations~\cite{gmeiner_exploring_2025}) ranged from one to five per session ($M=3.0, SD=1.6$).
Messages primarily helped people improve the part's obstacle geometry by considering additional factors such as bolt and tool clearances (Table \ref{tab:ch5-results-summary-table}). 
Design quality improved slightly compared to the baseline ($M=1.6$, $SD=0.9$ vs.\ $M=1.0$, $SD=0.0$), but many participants in both groups still failed to specify the bracket's structural loads and obstacle geometry adequately, leading to parts that did not pass all design criteria.  

In sum, while the system’s performance was weaker than hoped, the first Hybrid WoZ cycle provided critical methodological insight: live wizarding exposed misclassifications that were invisible in offline testing, and the structured record of message denials pointed us toward concrete refinements in the prompt design. 
These lessons directly informed the next iteration, where we began separating task-phase classification and message generation into distinct prompts.

\subsubsection{\textbf{Refining Prompt Structure by Separating Classification from Generation}}  
\label{findings-woz-prompt-refinements}

Insights from the first Hybrid WoZ cycle underscored that misclassifications of user activity were a central bottleneck. 
Methodologically, this highlighted the need to disentangle different reasoning steps within the prompting pipeline. 
In response, we adopted a prompt chaining approach~\cite{wu_ai_2022}, restructuring the system into two stages: one prompt dedicated to \textit{task-phase classification} and another for \textit{message generation} (Figure \ref{fig:ch5-prompt-evolution}). 
This separation allowed us to assess the model’s classification accuracy independently from the usefulness of its generated messages, and gave the wizard explicit points of intervention.  

In practice, the classification prompt first inferred whether the user was in the task phase of, for example, specifying constraints, defining geometry, or evaluating outcomes. 
The message-generation prompt then produced a support message tailored to that phase and to the chosen message type (question, suggestion, or software tip).  
The wizard could inspect each classification, override it when it did not fit the observed activity, and only then request a message. 
This setup retained human control while providing a clearer view into the model’s reasoning process.  

Before re-running Hybrid WoZ sessions, we used \textit{Counterfactual Video Replay Prompting} to refine the classification instructions. 
Replays across prior recordings exposed recurring errors, which we addressed by expanding the classification prompt with more explicit guidance and mistake–correction patterns. 
For example, commit \texttt{13b9d5a} [\textit{Task phase classification: More specific explanation of challenges and how to recognize these}] (Appendix Table \ref{tab:git-commits}) introduced detailed criteria for recognizing challenges in load specification, obstacle geometry, manufacturability, simulation, and outcome evaluation.  
These incremental changes, documented through Git commits, allowed us to track how prompt revisions improved the model’s ability to classify phases correctly.  

With this refined pipeline in place—classification first, message generation second, and wizard oversight at each step—we proceeded to a second cycle of Hybrid WoZ sessions to validate the new setup in real time.

\subsubsection{\textbf{Validating Improvements through a Second Hybrid WoZ Cycle}}  
\label{findings-woz-2}

The second Hybrid Wizard-of-Oz cycle provided an opportunity to test the revised prompting pipeline in live sessions. 
Methodologically, separating task-phase classification from message generation proved valuable: it surfaced the model’s reasoning more transparently, allowed the wizard to intervene at each stage, and enabled us to evaluate classification accuracy and message usefulness independently.  
In practice, the wizard overrode roughly half of the classifications (by choosing a different task phase for message generation than the one suggested by the model), but this was a relatively lightweight intervention with outsized effects: once an incorrect task phase was corrected by the wizard, the downstream message was still generated by the model. 
This direct control prevented missteps from propagating, created a clear record of where the model continued to struggle, and highlighted how stabilizing classification greatly improved the usefulness of the AI-generated messages.

Compared to the first iteration, the generated support messages were noticeably less repetitive and more targeted. 
By exposing classification results in the interface and allowing real-time overrides, the generated messages fit more closely with unfolding user activity. 
Overall, this combination of transparency and control not only improved the immediate quality of support messages but also provided valuable insights for evaluating our classification and message generation prompts in isolation. This made it clear that message generation produced satisfactory results once the correct task phase was specified.

At the outcome level, designers in this cycle achieved significantly higher-quality results. 
All participants created brackets that passed structural analysis, and most designs also met clearance and space constraints. 
The average outcome score rose to $M=4.4$ ($SD=0.5$), compared to $M=1.6$ ($SD=0.9$) in the first cycle (Table \ref{tab:ch5-results-summary-table}). 
Support messages more effectively guided designers to specify relevant inputs for the generative solver, resulting in more robust designs.  

In sum, the second Hybrid WoZ cycle showed that the refined pipeline offered both methodological and practical benefits. 
It provided a clear framework for diagnosing and fixing errors, which enabled more reliable delivery of useful support messages and resulted in measurable improvements in design outcomes.  
While task-phase classification still required further refinement, this stage of prototyping demonstrated that the overall system could reliably support users and provided us with a strong foundation for advancing SocraBot toward an effective and usable real-time support agent.

\subsection{Summary of Prototyping Methods and Insights}

We encountered several prototyping challenges and addressed them through two complementary methods: 

\noindent First, in prototyping a multimodal real-time support agent on top of a pretrained VLM, we faced the dual \textbf{challenge of having only limited user data for evaluation while also lacking established evaluation benchmarks.}
We then initially wanted to \textbf{compare prompt effects in a tabular format} (inspired by benchmark-style approaches), \textbf{but were unable to assess the usefulness of model outputs} (proactive contextual support messages) due to missing the unfolding task context---such as what the user had already done and tried in the task session, what strategies they were pursuing, or what misconceptions they held.

To address this, \textbf{\textit{Counterfactual Video Replay Prompting} allowed us to immerse ourselves in recorded user scenarios and flexibly test how different prompt strategies would play out across varying user contexts.} 
This revealed that a single, lengthy system prompt was too complex for the model to follow (which is generally a known limitation of language models, and figuring out \textit{when} and \textit{how} to split up prompts is part of the iterative trial-and-error process). 
\textbf{We therefore restructured the system prompt into smaller, specialized prompts} for different support types \textit{(reflective questions, design suggestions, and software tips)}, which produced more controlled and targeted support behavior.
    
However, counterfactual replay prompting \textbf{did not allow us to explore system behavior in entirely new user situations}, since the actions in the recordings were fixed. 
To address this limitation, we \textbf{introduced an adapted \textit{Hybrid Wizard-of-Oz} method} that enabled evaluation of prompt strategies during live user sessions. 
Here, real-time interaction revealed frequent misclassifications of user activities by the model, which led to irrelevant or frustrating support messages. 
Based on these observations, we \textbf{split the prompting pipeline into two stages—task-phase classification and message generation—and added mechanisms for manual overrides}, giving the facilitator more control to steer and document the model’s behavior during sessions. 

We then cycled back to counterfactual replay prompting, using it to refine the classification prompt and test corrections against prior recordings. 
This helped us diagnose and further reduce recurring misclassification errors. 
A subsequent round of hybrid Wizard-of-Oz sessions validated these refinements in real time: messages became less repetitive, more targeted, and better aligned with user needs, which led to improved design outcomes.  
In this way, the two methods complemented one another—counterfactual replays supported rapid, immersive testing of prompt variations, while hybrid Wizard-of-Oz sessions provided experiential validation in live contexts. 
Together, they enabled us to iteratively shape the prompting strategy as a foundation for a more automated version of SocraBot.

\section{The Counterfactual Replay Prompt Evaluation Toolkit}

Beyond methodological insights, we contribute an open-source toolkit that supports one key element of our approach: counterfactual replay prompting.
The toolkit is a self-contained application that enables HCI researchers and practitioners to rapidly prototype multimodal prompt-based GenAI model behaviors by replaying recorded user sessions and generating model outputs at arbitrary points in time (see Figure \ref{fig:teaser} and the video figure~\footnote{\url{https://youtu.be/jZCNyq1rzuM}.}). 
We refactored the internal tools developed during the SocraBot project into a reusable package with clear documentation, modular components, and a streamlined interface, making it easier for others to replicate our approach, adapt it to different domains, and extend its functionality. 
In this section, we illustrate a typical workflow, provide an overview of the toolkit's architecture, and highlight opportunities for integration and extension.  
The toolkit is released as open-source software under the MIT license, with the source code available in a public Git repository\footnote{An anonymized link to the repository is available at \textbf{\url{https://anonymous.4open.science/r/replay-prompt-eval-7C5D/}}. 
}.

\subsection{Workflow Scenario}

A typical usage scenario unfolds as follows:
\begin{enumerate}
    \item The user starts by loading a session recording by selecting the corresponding file from the session dropdown menu. 
    \item They navigate through the replay (video and transcript) to a moment of interest. 
    \item By clicking the message generation button, an LLM response is requested for the selected context. 
    Once generated, the new message appears in the interface. 
    \item The user evaluates the message, optionally assigning a five-point rating and leaving a comment. 
    \item They can iterate by adjusting the system prompt, replaying the same session, or testing prompts across different sessions. 
\end{enumerate}

In addition, the application includes a \textbf{coding view} that lets users label and annotate generated messages.
This view provides an overview of all outputs for a given session, along with associated ratings and comments, and allows jumping back to the corresponding video moment.
Updated annotations are stored automatically in the backend database.

\subsection{Architecture and Components}

\paragraph{\textbf{Frontend (Replay \& Interaction).}}
\begin{itemize}
    \item \textbf{Replay interface:} A web-based video player with synchronized transcripts. 
    Users can load recordings (MP4 + transcript/SRT) through a session dropdown and then play, scrub, or pause the recording. 
    \item \textbf{System prompt editor:} The current system prompt can be edited in a dedicated text field. 
    \item \textbf{Message generation controls:} A button allows users to trigger LLM responses for the selected context. Generated messages are displayed directly in the interface. 
    \item \textbf{Evaluation interface:} Each generated message can be optionally rated with a five-point rating widget and annotated with a free-text comment, which are stored in the backend database. 
\end{itemize}

\paragraph{\textbf{Backend.}}
\begin{itemize}
    \item \textbf{Media management:} Users drop video files (MP4) and transcripts (SRT) into a designated backend \texttt{media} folder, from which the frontend loads them dynamically. 
    \item \textbf{Prompt assembly pipeline:} The backend handles the assembling of prompt and context from the video frames and transcript, and issues LLM API inference calls.
    \item \textbf{Data persistence:} Ratings and comments are stored in a lightweight file-based JSON database. 
    \item \textbf{Server implementation:} Implemented as a Node.js service with distinct API endpoints 
\end{itemize}

\subsection{Extensibility and Integration}

The toolkit is implemented as a modular server-client web application that allows components to be adapted or replaced, such as swapping the JSON store for a database or adding custom data processing pipelines.
It provides basic extension points that make it possible to connect with hybrid Wizard-of-Oz workflows.

Out of the box, the toolkit supports counterfactual replay with video recordings; for live hybrid WoZ use, users need to link the backend to their own applications (e.g., to a video call application's live video and transcription stream).
Since the database is JSON file-based, ratings and comments can be easily exported for further qualitative or quantitative analysis in other software.

\section{Discussion}

While prompt-based approaches have significantly reduced the effort needed to build GenAI prototypes, creating sophisticated user experiences with these models remains a substantial challenge. This work documents our first-hand experience implementing SocraBot, a proactive cognitive real-time support agent built atop a multimodal VLM, revealing the complex challenges that emerge during this process.

Drawing on prior formative work by Gmeiner et al.~\cite{gmeiner_exploring_2025}, which provided a comprehensive understanding of designers’ support needs and articulated concrete design principles, we nevertheless struggled (like many researchers and practitioners in other task domains, e.g.,~\cite{zamfirescu-pereira_herding_2023, harisubramonyam_contentcentric_2024}) to translate this understanding into effective prompting strategies.
The core difficulty stemmed from a fundamental uncertainty: 
For such a domain-specific task without existing benchmarks, we had no reliable way to evaluate whether our prompting approach would deliver the intended experience when confronted with the nuanced complexity of real-world and real-time user interactions. 
Our initial reliance on semi-automated tabular evaluations proved inadequate for capturing the contextual richness of such time-based multimodal interactions.
This uncertainty drove us to develop a custom prototyping and evaluation pipeline combining Counterfactual Replay Prompting with a Hybrid Wizard-of-Oz approach. 
These methods enabled us to refine SocraBot's prompting structure iteratively while maintaining a direct connection to authentic user contexts. 
By immersing ourselves in realistic usage scenarios, we could directly observe how different prompting strategies affected system behavior across varying situations---insights that traditional evaluation methods would have failed to surface at this early prototyping stage.

In the following sections, we discuss the implications of these novel approaches for prototyping multimodal real-time GenAI agent experiences. 
We examine how they address complexity challenges in prompt engineering, complement existing prototyping methods, and support a more designerly approach to creating GenAI-based systems that meet user needs in complex, real-world contexts.

\subsection{Taming Complexity with Iterative Prompt Decomposition Through Interactive Human-in-the-loop Steering and Evaluation }

This case study highlights how evaluating GenAI model performance for novel multimodal tasks without established test data and benchmarks feels like navigating blindfolded. 
In these cases, practitioners typically resort to iterative trial-and-error testing when working with prompt-based models~\cite{zamfirescu-pereira_why_2023}. 
Currently existing ad-hoc prompt evaluation tools largely focus on text generation~\cite{kim_evallm_2024, arawjo_chainforge_2024}, falling short in handling the multimodal and time-based data needed for evaluating proactive real-time systems like SocraBot. 
This limitation becomes more evident when working with emerging multimodal capabilities that combine visual, verbal, and contextual interactions. 
Overall, our research demonstrates how interactive and experiential evaluation methods such as \textit{Counterfactual Replay Prompting} and \textit{Hybrid Wizard-of-Oz} bridge the gap between blind trial-and-error and rigid automated benchmark testing. 

These approaches fostered deeper problem understanding early in the development process by letting us observe how model outputs played out in realistic task contexts.  
Counterfactual replays, for example, revealed how suggestions that looked linguistically sound in a table often failed when judged against the unfolding user activity (Section \ref{findings-rq1}).  
Hybrid Wizard-of-Oz sessions, in turn, exposed how misclassifications of user context (such as their current task phase) could cascade into ineffective support behaviors,\footnote{Similar cascading errors have been observed in chain-of-thought and reasoning pipelines, where early missteps compound into larger failures~\cite{zhao_uncertainty_2025,you_probabilistic_2025}.} directly motivating our restructuring of the prompt architecture into smaller, specialized modules (Sections \ref{findings-woz-1}--\ref{findings-woz-2}).

By \textbf{offering a gradual path toward automation}, these methods allow practitioners to begin with highly interactive and situated evaluation before transitioning to more automated testing as their understanding improves. 
Rather than replacing benchmark testing, the methods we share in this paper serve as a crucial bridge for emerging models and task domains where test data is scarce and effective prompting strategies remain unclear. 
Such methods also provide experiential insights into model behavior that help shape what \textit{kinds of data} need to be collected for more systematic evaluation in the future. 

Based on this case study, such methods prove especially valuable in taming complexity during the early stages of developing multimodal real-time agent systems.  
They enabled us to decompose complex prompt structures into smaller units while still maintaining a holistic view of how these modules function together in a pipeline.  
For instance, replaying full sessions revealed a skewed distribution of message types. 
This insight, visible only when prompts were tested against unfolding user activity, guided us to split these categories into separate prompt modules and iteratively evaluate their balance within the pipeline (Section \ref{ch5:findings2}).

\subsection{Supporting Designerly and User-centered Prompt Engineering Approaches}

The way we employed these prototyping methods aimed to support a \textit{holistic practice} of prompt engineering.
Designers must attend not only to users—whose goals, skills, and practices vary widely—but also to stochastic ML models, whose behaviors are inconsistent and require careful probing.
User-centered design, in this light, becomes a mediating practice between two sources of uncertainty: people and models.
The designer’s role is to continually negotiate this space, refining prompts so that model behavior aligns with evolving user workflows.

Immersing ourselves in concrete user situations—through video replay or during hybrid Wizard of Oz sessions—allowed us to develop a comprehensive, contextual understanding of how to structure and chain prompt instructions without requiring massive upfront data collection.
Though time-intensive, the hybrid Wizard of Oz sessions provided crucial insights into model behavior across various situations (see Sections \ref{findings-woz-1} and \ref{findings-woz-2}). 
This helped us develop a tacit understanding of how the model responded to different inputs, which directly informed our prompt structure adjustments.

A valuable insight emerged during hybrid Wizard of Oz testing when we could observe and override the model's misclassifications in real time (see \ref{findings-woz-2}). 
Seeing when the model made incorrect classifications in the first chain element (classifying the task phase) and then overriding these decisions for subsequent message generation provided an effective mechanism to assess individual prompt effects and their interplay. 
Iteratively moving back and forth between counterfactual replays and hybrid WoZ cycles also proved powerful: replay sessions allowed us to capture more data, experiment rapidly with new prompt ideas, and refine them in a low-stakes setting, while hybrid WoZ sessions tested these refinements in live interaction. 
This cyclical process allowed us to bootstrap the agent while being embedded in the interaction itself, offering a unique form of experiential feedback that is rarely available through conventional evaluation.

Overall, the proposed value of these experiential prompting methods lies in how they support reflective practice in prompt engineering. 
Unlike traditional evaluation approaches focused on technical metrics, these methods emphasize the practitioner's developing relationship with the model, revealing how its behaviors respond to various inputs and contexts. 
This approach recognizes that effective prompt engineering involves both technical skill and ``designerly'' intuition developed through immersive observation and interaction. 
By supporting this reflective dialogue between designer and system, these methods help practitioners develop the tacit knowledge needed to create more responsive, contextually aware GenAI experiences that genuinely address user needs rather than merely optimizing for technical performance metrics.

Such intuition-building prototyping practice is particularly important in the context of cognitive assistant design, where performance metrics are difficult to obtain: support unfolds over long-running tasks, and clear signals of effectiveness may only appear after extended interaction. 
Unlike question-answering systems, where every exchange provides measurable feedback, assistants for complex design work offer fewer and slower signals, making experiential, in-situ evaluation a crucial part of the prototyping process.

\subsection{Future Synergies with Existing Methods and Tools}

The case study, prototyping process, and toolkit described in this paper build upon existing methods such as data replays, hybrid Wizard of Oz, and prompt-evaluation frameworks. While surfacing insights about the specific prototyping process within one project and domain (design support agent), the case study also revealed exciting opportunities for integration with existing methods and other task domains.

For example, a promising avenue involves incorporating Counterfactual Video Replay Prompting into existing prompt evaluation tools. Current evaluation frameworks (e.g.,~\cite{arawjo_chainforge_2024,kim_evallm_2024}) provide rich interfaces for text-based prompt assessment but lack support for multimodal, time-based data. 
By integrating video-based replay capabilities, such tools could be expanded for evaluating multimodal GenAI experiences.
Similarly, model pipeline editors (e.g.,~\cite{lin_jigsaw_2024, comfyanonymous_comfyanonymous_2025}) could benefit from incorporating replay-based evaluation methods, enabling practitioners to test how different model pipelines respond to authentic multimodal user scenarios rather than hypothetical examples. 
This integration would provide more nuanced insights into how prompt chains perform across varying contexts and user behaviors.
The Hybrid Wizard-of-Oz approach could be extended beyond video-based interactions to encompass emerging spatial computing platforms such as virtual and augmented reality. 
As multimodal GenAI models increasingly power spatial immersive experiences, these methods could therefore help designers evaluate and refine prompt strategies for novel spatial interaction tasks where contextual understanding becomes even more complex.

Another promising opportunity lies in combining evaluation with data labeling. 
While the methods presented in this paper primarily focus on refining prompt strategies through observation and intervention, they could also be extended to capture human interventions in steering model behavior. 
These intervention points might represent valuable training data for improving the model performance itself. 
By documenting when and why human overrides occur, we could build feedback loops that allow GenAI models to learn from these corrections over time, gradually reducing the need for human mediation. 
In addition, these processes could be combined with reflective prompt generation strategies, such as GEPA~\cite{agrawal_gepa_2025}, which enable AI systems to iteratively refine their own prompts through natural language reflection. 
While such reflective optimization holds promise for reducing human effort in prompt engineering, our findings highlight the continued importance of designer-led evaluation to ensure that evolving prompts remain aligned with real user needs and contextual constraints.

In conclusion, all these potential synergies illustrate how methods such as Counterfactual Replay Prompting and Hybrid Wizard of Oz can complement and enhance existing prototyping approaches, creating more comprehensive evaluation ecosystems that address the unique challenges of multimodal, context-sensitive, real-time GenAI applications across diverse interaction paradigms and task domains. 
The open-source toolkit we developed provides a practical foundation for such integrations: it can be extended with modules for integrating replay-based evaluation and human-in-the-loop live steering and labeling into other prompt evaluation pipelines, and adapted to domains beyond cognitive design assistants. 
In doing so, it offers a bridge between research prototypes and broader community use, supporting both academic experimentation and practitioner-driven tool development.

\subsection{Limitations}

While the methods described in this work offer promising advantages for prototyping multimodal GenAI experiences, several limitations stand out.  
First, this research centers on a single case study in a specific domain, and further research is required to validate its generalizability across other GenAI applications. 
Second, these approaches face scalability challenges due to their reliance on human oversight, making large-scale simultaneous evaluations impractical. 
However, their primary aim is not scale but deep contextual insight during early-stage prompt development.
Third, the Hybrid WoZ approach introduces potential wizard bias, as human judgments on \textit{when} to intervene can influence outcomes and prompt iterations. 
Yet such intervention is not merely a limitation but also a vital part of the learning and refinement process in early-stage prototyping—consistent with the “fake it to make it” tradition in HCI, where human scaffolding helps probe what future automation should deliver~\cite{stadler_fake_2023a}. 
The remaining challenge lies in automating intervention timing: inferring when proactive support should occur is still an open problem in HCI, and remains particularly difficult to standardize or automate.

Reproducibility is also a concern; human involvement (especially when overriding LLM outputs) makes it challenging to recreate consistent evaluation conditions, complicating the systematic comparison of prompt strategies across evaluators. 
These challenges can be mitigated through precautions such as clearly documenting decision criteria, standardizing logging practices, and, in team settings, conducting cross-rater discussions.  
Ultimately, these methods are intended as tools for experiential understanding, which ideally will lay the groundwork for more systematic and automated testing in later stages.
Finally, Counterfactual Replay Prompting depends on pre-recorded sessions, which may not reflect the full range of user behavior, potentially leading to prompt strategies tuned to narrow interaction patterns rather than real-world variability.

\section{Conclusion}

Recent advances in multimodal GenAI open up possibilities for building proactive, context-aware agents that can follow users’ activities and provide real-time assistance. 
Yet, prototyping such systems remains difficult, as existing prompt engineering and evaluation methods fall short in addressing the situational complexity and immediacy of real-world workflows. 
Through our case study of building \textit{SocraBot}, we developed a user-centered prototyping pipeline comprising \textit{Counterfactual Video Replay Prompting} and a \textit{Hybrid Wizard-of-Oz} approach. 
These complementary methods allowed us to iteratively decompose and refine prompts, evaluate model outputs across both recorded and live contexts, and better align system behaviors with users’ unfolding activities.   
Our experiences highlight how counterfactual replay and hybrid Wizard-of-Oz strategies can help HCI researchers and practitioners tame the uncertainty of early-stage GenAI prototyping by grounding evaluation in experiential, context-rich user data. 
Beyond our specific case study, we contribute the open-source \textit{Counterfactual Replay Toolkit}, enabling others to rapidly test and adapt multimodal prompting strategies for their own domains. 
Together, these contributions open new pathways for building and evaluating proactive, context-aware, real-time AI agents, and point to future opportunities for developing more robust methods and infrastructures that support the iterative, user-centered prototyping of emerging GenAI systems. Looking ahead, we see this work as a step toward an expanded methodological repertoire for HCI+AI that enables researchers and practitioners to more effectively study and shape the evolving role of GenAI real-time agents in real-world human workflows.

\begin{acks}
We thank all study participants and the research assistants Anna Xu and Claire Malella for supporting this work.
This material is based upon work supported by the National Science Foundation under Grant No. \#2118924 Supporting Designers in Learning to Co-create with AI for Complex Computational Design Tasks.
\end{acks}


\appendix

\newpage
\onecolumn

\section{Additional Materials}

\begin{table}[h]
\caption{Overview of study participants. 
}
\begin{tabular}{llclllllll}
\toprule
\textbf{ID } & 
\textbf{Group}  & 
\textbf{Age} & 
\textbf{Role}                                &
\begin{tabular}[c]{@{}l@{}}\textbf{MechDes} \\ \textbf{Exp.} \\ \textbf{Years}\end{tabular} & 
\begin{tabular}[c]{@{}l@{}}\textbf{Indus.} \\ \textbf{Exp.} \\ \textbf{Years}\end{tabular} & 
\begin{tabular}[c]{@{}l@{}}\textbf{CAD} \\ \textbf{Exp.} \\ \textbf{Years}\end{tabular}& 
\begin{tabular}[c]{@{}l@{}}\textbf{FEA } \\ \textbf{Prof.}\end{tabular}&  
\begin{tabular}[c]{@{}l@{}}\textbf{DFM } \\ \textbf{Prof.}\end{tabular}\\

\midrule
A1 & HyWoZ Cycle 1 & 20 & Student, BS MechE  & 3 – 5   & 1 – 2 & 5+    & 3 & 2 \\
A2 & HyWoZ Cycle 1 & 25 & Student, MA MechE  & 3 – 5   & 1 – 2 & 2 – 4 & 5 & 6 \\
A3 & HyWoZ Cycle 1 & 21 & Student, BS MechE  & 3  –  5 & None  & 2 – 5 & 4 & 5 \\
A4 & HyWoZ Cycle 1 & 22 & Student, MS MechE  & 6 – 10  & 3 – 5 & 5+    & 6 & 5 \\
A5 & HyWoZ Cycle 1 & 22 & Student, MS MechE  & 1 – 2   & 0     & 2 – 4 & 3 & 2 \\
B1 & HyWoZ Cycle 2 & 19 & Student, BS AerospE   & 6 – 10  & None  & 5+    & 4 & 3 \\
B2 & HyWoZ Cycle 2 & 21 & Student, BS MechE  & 3 – 5   & 1 – 2 & 5+    & 3 & 5 \\
B3 & HyWoZ Cycle 2 & 20 & Student, BS MechE  & 3 – 5   & 3 – 5 & 2 – 4 & 5 & 7 \\
B4 & HyWoZ Cycle 2 & 19 & Student, BS MechE  & 1 – 2   & None  & 2 – 4 & 1 & 1 \\
B5 & HyWoZ Cycle 2 & 43 & Engineer                            & 10+     & 10+   & 5+    & 4 & 7 \\             
\bottomrule
\end{tabular}
\label{tab:ch5-appendix-participants}
\Description{The table presents an overview of study participants, detailing their identification numbers (ID), group affiliations, ages, and roles. It further includes their experience in mechanical design, industrial environments, and CAD software, as well as their proficiency ratings in FEA and DFM. Each row corresponds to an individual participant with specific values for the listed categories.}
\end{table}

\begin{table}[h]
\caption{Selected Git commits specifically related to iterative refinement of GPT system prompts (not a record of all commits).}
\label{tab:git-commits}
\Description{The table displays a list of selected Git commits, with two columns: "Commit" and "Message." Each row contains a unique commit identifier alongside a brief description of the changes made in that commit, focusing on the iterative refinement of GPT system prompts. The messages detail actions taken, such as adding support strategies and refining message descriptions.}
\centering
\begin{tabularx}{\linewidth}{l X}
\toprule
\textbf{Commit} & \textbf{Message} \\
\midrule
645a0bf & Initial commit   \\
14a39b0 & Prompt engineering \&  front end fixes \\
e949a69 & Added "reason for denying message" to prompt   \\
9d62ddc & Added additional support strategies to the prompt   \\
bac875d & Refined support messages description   \\
1cb8054 & Prompt message generation (time stamp fix)   \\
53b87d0 & Software tip type and planning phaser   \\
15d197d & Prompts separated   \\
3943d7e & Changes WOZ to include phase radio group and updated prompt to respect phase   \\
6840f8e & Added task phase classification mechanism   \\
13b9d5a & Task phase classification: More specific explanation of challenges and how to recognize these.   \\
\bottomrule
\end{tabularx}
\end{table}

\end{document}